\documentclass[12pt]{spieman}  % 12pt font required by SPIE;
\usepackage{amsmath,amsfonts,amssymb}
\usepackage{graphicx}
\usepackage{setspace}
\usepackage{tocloft}
\usepackage{ulem}

\title{Directly Characterizing the Coherence of Quantum Detectors by Sequential Measurement}

\author[a,b$\dagger$]{Liang Xu}
\author[a,c$\dagger$]{Huichao Xu}
\author[a]{Jie Xie}
\author[a]{Hui Li}
\author[a]{Lin Zhou}
\author[a]{Feixiang Xu}
\author[a*]{Lijian Zhang}

\affil[a]{National Laboratory of Solid State Microstructures, Key Laboratory of Intelligent Optical Sensing and Manipulation, College of Engineering and Applied Sciences, and Collaborative Innovation Center of Advanced Microstructures, Nanjing University, Nanjing 210046, China}
\affil[b]{Research Center for Quantum Sensing, Zhejiang Lab, Hangzhou 310000, China}
\affil[c]{Purple Mountain Laboratories, Nanjing, Jiangsu 211111, China}

\cftpagenumbersoff{figure}
\cftpagenumbersoff{table} 
\begin{document} 
\maketitle

\begin{abstract}
The quantum properties of quantum measurements are indispensable resources in quantum information processing and have drawn extensive research interest. The conventional approach to reveal the quantum properties relies on the reconstruction of the entire measurement operators by quantum detector tomography. However, many specific properties can be determined by a part of matrix entries of the measurement operators, which provides us the possibility to simplify the process of property characterization. Here, we propose a general framework to directly obtain individual matrix entries of the measurement operators by sequentially measuring two non-compatible observables. This method allows us to circumvent the complete tomography of the quantum measurement and extract the useful information for our purpose. We experimentally implement this scheme to monitor the coherent evolution of a general quantum measurement by determining the off-diagonal matrix entries. The investigation of the measurement precision indicates the good feasibility of our protocol to the arbitrary quantum measurements. Our results pave the way for revealing the quantum properties of quantum measurements by selectively determining the matrix entries of the measurement operators.
\end{abstract}

% Include a list of up to six keywords after the abstract
\keywords{direct tomography, quantum measurement, weak measurement, sequential measurement, coherence}

% Include email contact information for corresponding author
{\noindent \footnotesize\textbf{*}Lijian Zhang,  \linkable{lijian.zhang@nju.edu.cn} }

\begin{spacing}{2}   % use double spacing for rest of manuscript

\section{Introduction}
The quantum properties of quantum measurements have been widely regarded as essential quantum resources for the preparation of quantum states \cite{ourjoumtsev2007generation, bimbard2010quantum, ulanov2016loss}, achieving the advantages of quantum technologies \cite{higgins2007entanglement, knill2001scheme, resch2007time, kok2007linear} as well as the study of fundamental quantum theories \cite{zhang2019experimental, li2017experimental, frustaglia2016classical, markiewicz2019contextuality, berg2015classically, guzman2016demonstration, ndagano2017characterizing, goldberg2020extremal}. The time-reversal approach allows to investigate the properties of quantum measurements qualitatively from the perspective of quantum states \cite{barnett2000retrodiction, pegg2002quantum, amri2011characterizing}. In addition, the quantum resource theories to properly quantify the quantum properties of quantum measurements has been developed very recently \cite{theurer2019quantifying, cimini2019measuring, bischof2019resource, guff2021resource}, and has been applied to investigate an important quantum property, the coherence, of quantum-optical detectors \cite{xu2020experimental}. Thus, developing efficient approaches to characterize the quantum properties of quantum measurements is important for both the fundamental investigations and practical applications.

A general quantum measurement, and all its properties, can be completely determined by the positive operator-valued measure (POVM) $\{\hat{\Pi}_l\}$, in which the element $\hat{\Pi}_l$ denotes the measurement operator corresponding to the outcome $l$. Several approaches have been developed to determine the unknown POVM \cite{lundeen2009tomography, d2004quantum, luis1999complete, fiuravsek2001maximum}, of which the most representative one is quantum detector tomography (QDT) \cite{lundeen2009tomography}. In QDT, a set of probe states $\{\rho^{(m)}\}$ are prepared to input the unknown measurement apparatus and the probability of obtaining the outcome $l$ is given by $p_l^{(m)} = \text{Tr}(\rho^{(m)}\hat{\Pi}_l)$. Provided that the input states are informationally complete for the tomography, the POVM $\{\hat{\Pi}_l\}$ can be reconstructed by minimizing the gap between the theoretical calculation and the experimental results. To date, QDT has achieved great success in characterizing a variety of quantum detectors, including avalanche photodiodes \cite{d2011quantum}, time-multiplexed photon-number-resolving detectors \cite{lundeen2009tomography, feito2009measuring, coldenstrodt2009proposed}, transition edge sensors \cite{Brida_2012}, and superconducting nanowire detectors \cite{akhlaghi2011nonlinearity}. As the quantum detectors become increasingly complicated, the standard QDT is confronted with the experimental and computational challenges, which prompts the exploration of some helpful shortcuts. For example, the determination of a few key parameters that describe the quantum detectors allows to largely reduce the characterization complexity \cite{worsley2009absolute}. The quantum detectors can also be self-tested with certain quantum states in the absence of the prior knowledge of the apparatus \cite{mayers1998quantum, gomez2016device, zhang2018experimentally, tavakoli2018self}. The emerging data-pattern approach realizes the operational tomography of quantum states through fitting the detector response, which is robust to imperfections of the experimental setup \cite{vrehavcek2010operational, mogilevtsev2013data}.

Though QDT is a generic protocol to acquire the entire measurement operators, it does not have the direct access to the single matrix entries of the measurement operator. The complexity of the reconstruction algorithm in
QDT increases dramatically with the increase of the dimensional of the quantum system. Typically, tomography of the full measurement operators is considered as the prerequisite for characterizing the properties of quantum measurements \cite{xu2020experimental}. However, in some situations, the complete determination of the measurement operators is not necessary to fulfil specific tasks, which makes it possible to simplify the characterization process. For example, if the input state is known to lie in the subspace of the quantum system, it only requires the corresponding matrix entries of the measurement operators to predict the probability of outcomes \cite{zhang2012mapping, renema2014experimental, feito2009measuring}. In particular, the coherence of a quantum measurement is largely determined by the off-diagonal matrix entries of its measurement operators in certain basis \cite{xu2020experimental}.

Recently, Lundeen et.al proposed a method to directly measure the probability amplitudes of the wavefunction using the formalism of the weak measurement and weak values \cite{lundeen2011direct}. This method, known as the direct quantum state tomography, opens up a new avenue for the quantum tomography technique. The direct tomography (DT) protocol has been extensively studied and the scope of its application is expanded to high-dimensional states \cite{malik2014direct, howland2014compressive, mirhosseini2014compressive, knarr2018compressive, shi2015scan, ogawa2019framework, ogawa2021direct, zhou2021direct, yang2020zonal}, mixed states \cite{salvail2013full, bamber2014observing, lundeen2012procedure, thekkadath2016direct, wu2013state} and entangled states \cite{ren2019efficient, pan2019direct}, quantum processes \cite{kim2018direct} and quantum measurements \cite{xu2021direct}. The development of the DT theory from the original weak-coupling approximation to the rigorous validation with the arbitrary coupling strength ensures the accuracy and simultaneously improves the precision \cite{zou2015direct, vallone2016strong, zhang2016coupling, zhu2016direct, denkmayr2017experimental, calderaro2018direct, zhu2019hybrid, zhang2020direct}. Moreover, the direct state tomography allows to directly measure any single matrix entry of the density matrix, which has provided an exponential reduction of measurement complexity compared to the standard quantum state tomography in determining a sparse multiparticle state \cite{chen2021directly, bamber2014observing, lundeen2012procedure, thekkadath2016direct, wu2013state}. Recent work has extended the idea to realize the direct characterization of the full measurement operators based on weak values, showing the potential advantages over QDT in the operational and computational complexity \cite{xu2020direct}. In view of the unique advantages of the DT, it is expected that the generalization of the DT scheme to directly characterizing the matrix entries of the measurement operators allows to extract the properties of the quantum measurement in a more efficient way.

In this paper, we propose a framework to directly characterize the individual matrix entries of the measurement operators by sequentially measuring two non-compatible observables with two independent meter states. In the following, the unknown quantum detector performs measurement on the quantum system. The specific matrix entry of the measurement operator can be extracted from the collective measurements on the meter states when the corresponding outcomes of the quantum detector are obtained. Our procedure is rigorously valid with the arbitrary non-zero coupling strength. The investigations of the measurement precision indicate the good feasibility of our scheme to characterize arbitrary quantum measurement. We experimentally demonstrate our protocol to monitor the evolution of coherence of the quantum measurement in two different situations, the dephasing and the phase rotation, by characterizing the associated off-diagonal matrix entries. Our results show the great potential of the DT in capturing the quantum properties of the quantum measurement through partial determination of the measurement operators.

\section{Theoretical Framework}
\label{sect:theory}
\subsection{Directly Determining the Matrix Entries of the Measurement Operators}
\label{sect:theory1}
\begin{figure}
\begin{center}
\begin{tabular}{c}
\includegraphics[height=4.3cm]{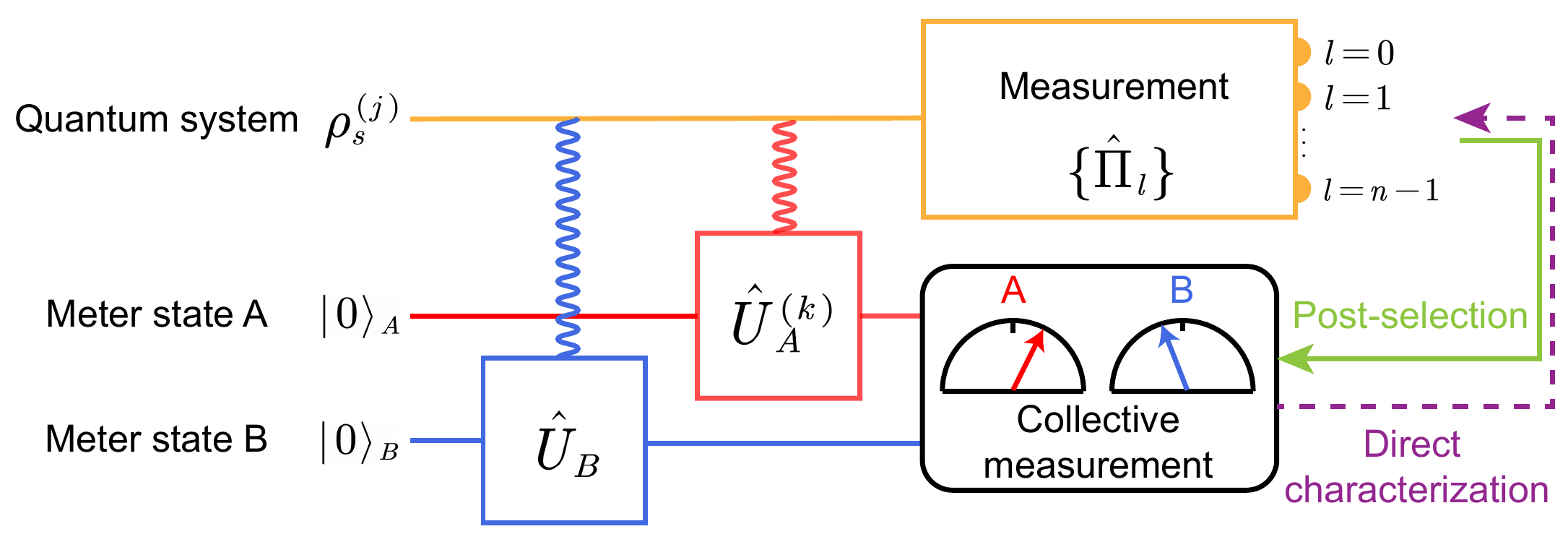}
\end{tabular}
\end{center}
\caption 
{ \label{schematic}
The schematic diagram for direct characterization of the matrix entries of the POVM $\{\hat{\Pi}_l\}$. } 
\end{figure}

The schematic diagram of directly measuring the matrix entries of the POVM is shown in Fig. \ref{schematic}. We represent the POVM $\{\hat{\Pi}_l\}$ acting on the $d$-dimensional quantum system (QS) with the orthogonal basis $\{|a_j\rangle\}$ ($\mathcal{A}$) and the matrix entry of the measurement operator $\hat{\Pi}_l$ is given by $E_{a_ja_k}^{(l)} = \langle a_j|\hat{\Pi}_l|a_k\rangle$. If $j=k$, $E_{a_ja_k}^{(l)}$ corresponds to the diagonal matrix entry which can be easily determined by inputting pre-selected QS state $\rho_s^{(j)} = |a_j\rangle\langle a_j|$ to the quantum detector and collecting the probability $p_l = \langle a_j|\hat{\Pi}_l|a_j \rangle$ of obtaining the outcome $l$. By contrast, the off-diagonal matrix entry $E_{a_ja_k}^{(l)}$ ($j\neq k$), generally a complex number, is related to the coherence of the operator and usually indirectly reconstructed in the conventional QDT. In order to directly measure $E_{a_ja_k}^{(l)}$ ($j\neq k$), we perform the sequential measurement of the observables $\hat{O}_B = \hat{I} - 2 |b_0\rangle\langle b_0|$ ($|b_0\rangle$ is a superposition state of all the base states in basis $\mathcal{A}$ with the equal probability amplitudes, i.e., $|b_0\rangle \propto \sum_j |a_j\rangle$) and $\hat{O}_A^{(k)} = \hat{I} - 2 |a_k\rangle\langle a_k|$ on the initial state $\rho_s^{(j)}$ with two independent two-dimensional meter states initialized as $|0\rangle_B$ and $|0\rangle_A$, respectively. The measurement of the observable $\hat{O}$ (generally referring to the observable $\hat{O}_B$ or $\hat{O}_A^{(k)}$) is implemented by coupling the QS with the meter state (MS) under the Hamiltonian $\hat{H} = g\delta(t-t_0)\hat{O}\otimes\hat{\sigma}_{y}$, in which $g$ is the coupling strength and $\hat{\sigma}_y = i(|1\rangle\langle 0|-|0\rangle\langle 1|)$ is the observable of the MS. Since the observables $\hat{O}_B$ and $\hat{O}_A^{(k)}$ do not commute, the measurement has to be performed in a particular order. 

The sequential measurement process can be described by the unitary evolution of the system-meter state $\rho_{sm} = \rho_s^{(j)}\otimes \rho_{m,B}\otimes \rho_{m,A}$ with the first transformation
\begin{equation}
\hat{U}_B = \exp(-ig_B\hat{O}_B\otimes \hat{\sigma}_{y,B}\otimes \hat{I}_A)
\end{equation}
and the second transformation
\begin{equation}
\hat{U}_A^{(k)} = \exp(-ig_A\hat{O}_A^{(k)}\otimes \hat{I}_B \otimes \hat{\sigma}_{y,A})
\end{equation}
leading to the joint state
\begin{equation}
\rho_J = \hat{U}_A^{(k)}\hat{U}_B \rho_{sm}\hat{U}_B^\dagger \hat{U}_A^{(k)\dagger}.
\end{equation}
Then, the unknown quantum detector to be characterized performs the post-selection measurement $\{\hat{\Pi}_l\}$ on the QS. Depending on the measurement outcome $l$, the surviving final meter state is given by $\rho_{m,A,B}^\prime = \text{Tr}_s(\hat{\Pi}_l\otimes \hat{I}_B \otimes \hat{I}_A \rho_J)/p_f$, in which $\text{Tr}_s(\cdot)$ denotes the partial trace operation on the QS and $p_f = \text{Tr}(\hat{\Pi}_l\otimes \hat{I}_B\otimes \hat{I}_A\rho_J)$ is the probability for getting the outcome $l$.

The matrix entry $E_{a_ja_k}^{(l)}$ is related to the average value of the observables $\hat{O}_B$ and $\hat{O}_A^{(k)}$ by
\begin{equation}
E_{a_ja_k}^{(l)} = \frac{d}{4} \text{Tr}\big[\hat{\Pi}_l(\hat{I}-\hat{O}_A^{(k)}) (\hat{I}-\hat{O}_B)\rho_s^{(j)}\big].
\label{eq:ele_relation}
\end{equation}
Both the observables $\hat{O}_B$ and $\hat{O}_A^{(k)}$ are designed to satisfy $\hat{O}^2 = \hat{I}$ so that the unitary is exactly expanded as $\hat{U} =\exp(-ig\hat{O}\otimes \hat{\sigma}_y)= \cos g\hat{I}\otimes \hat{I} - i\sin g \hat{O}\otimes\hat{\sigma}_{y}$. The right side of the Eq. \eqref{eq:ele_relation} can be extracted by the joint measurement of post-selected meter state $\rho_{m,A,B}^\prime$ with the observables
\begin{eqnarray}
\hat{P} &=& \sqrt{d}\Big(\frac{\hat{I}+\hat{\sigma}_z}{4\cos^2 g} -\frac{ \hat{\sigma}_x}{4\sin g\cos g}\Big), {}\nonumber\\
\hat{Q} &=& -\frac{\sqrt{d}\hat{\sigma}_y}{4\sin g\cos g}
\end{eqnarray}
each in the subsystems A and B. By defining the joint observables of meter states A and B as $\hat{R}_{B,A} = \hat{P}_B\hat{P}_A - \hat{Q}_B\hat{Q}_A$ and $\hat{T}_{B,A} = \hat{P}_B\hat{Q}_A + \hat{Q}_B\hat{P}_A$, we obtain the real and the imaginary parts of $E_{a_ja_k}^{(l)}$:
\begin{eqnarray}
\text{Re}[E_{a_ja_k}^{(l)}] &=& \text{Tr}(\hat{\Pi}_l\otimes \hat{R}_{B,A}\rho_J) {}\nonumber\\
\text{Im}[E_{a_ja_k}^{(l)}] &=& \text{Tr}(\hat{\Pi}_l\otimes \hat{T}_{B,A}\rho_J).
\label{R_I_measure}
\end{eqnarray}
Here, the subscripts of the coupling strength $g$ and the Pauli operators coincide with those of the operators $\hat{P}$ and $\hat{Q}$. For the sake of convenience, we take $g_A = g_B= g$ in the rest of this article.

\subsection{Precision analysis on directly characterizing the matrix entries of the measurement operators}
The accuracy and the precision are two essential indicators to evaluate a measurement scheme. There is no systematic errors in our protocol, since the derivation is rigorous for the arbitrary coupling strength $g$. According to the previous studies, the precision of the DT applied to the quantum states is sensitive to both the coupling strength and the unknown states \cite{PhysRevA.92.062133}. The increase of the coupling strength is beneficial to improve the precision \cite{vallone2016strong, zhang2016coupling, zhu2016direct, denkmayr2017experimental, calderaro2018direct, zhu2019hybrid, zhang2020direct}. When the unknown state approaches being orthogonal to the post-selected state, the DT protocol is prone to large statistical errors and therefore highly inefficient \cite{PhysRevA.92.062133, haapasalo2011weak}. Here, we theoretically investigate the precision of the DT protocol applied to the quantum measurement to verify the feasibility of our protocol.

Given that the real and the imaginary parts of the matrix entries are independently measured, we quantify the measurement precision with the total variance $\Delta ^2 E_{a_ka_j}^{(l)} =\Delta^2 \text{Re}[E_{a_ka_j}^{(l)}]+\Delta^2 \text{Im}[E_{a_ka_j}^{(l)}]$.
According to the Eq. \eqref{R_I_measure}, the variance can be derived by
\begin{equation}
\Delta ^2 E_{a_ka_j}^{(l)} = \langle \Delta^2 \hat{R}_{B,A}\rangle_f + \langle \Delta^2 \hat{T}_{B,A}\rangle_f,
\label{prec_ob}
\end{equation}
where $\langle \Delta ^2\hat{M}\rangle_f = \text{Tr}(\hat{\Pi}_l\hat{M}^2\rho_J) - [\text{Tr}(\hat{\Pi}_l\hat{M}\rho_J)]^2$. Since the operators $\hat{R}_{B,A}$ and $\hat{T}_{B,A}$ are usually hard to experimentally constructed, an alternative is to infer the expected values of $\hat{R}_{B,A}$ and $\hat{T}_{B,A}$ as well as their squares from the complete measurement results of the meter states B and A, each projected to the mutually unbiased bases (MUB), i.e., $\{|0\rangle,|1\rangle\},\{|+\rangle,|-\rangle\},\{|\circlearrowright \rangle,|\circlearrowleft \rangle \}$ with $|\pm\rangle  =(|0\rangle \pm |1\rangle)/\sqrt{2}$ and $|\circlearrowright\rangle, |\circlearrowleft\rangle = [ |0\rangle \pm i |1\rangle]/\sqrt{2}$. The obtained probability distribution is represented by $\{W_{mn}\}$, where $m$ and $n$ label the projective states $|m_B\rangle$ and $|m_A\rangle$ of the meter states B and A, respectively. The experimental variance can be obtained from $\{W_{mn}\}$ with the error transfer formula
\begin{equation}
\Delta^2 E_{a_ka_j}^{(l)} = \sum_{m,n} \Big|\frac{\partial E_{a_ka_j}^{(l)}}{\partial Pr_{mn}}\Big|^2 \delta^2 W_{mn}.
\label{Prec_err}
\end{equation}
Consider $N$ particles are used for one measurement of $W_{mn}$. The variance of the probability is approximated as $\delta^2 W_{mn}\approx W_{mn}/N$ in the large $N$ limit due to the Poissonian statistic.

\begin{figure}
\begin{center}
\begin{tabular}{c}
\includegraphics[height=5cm]{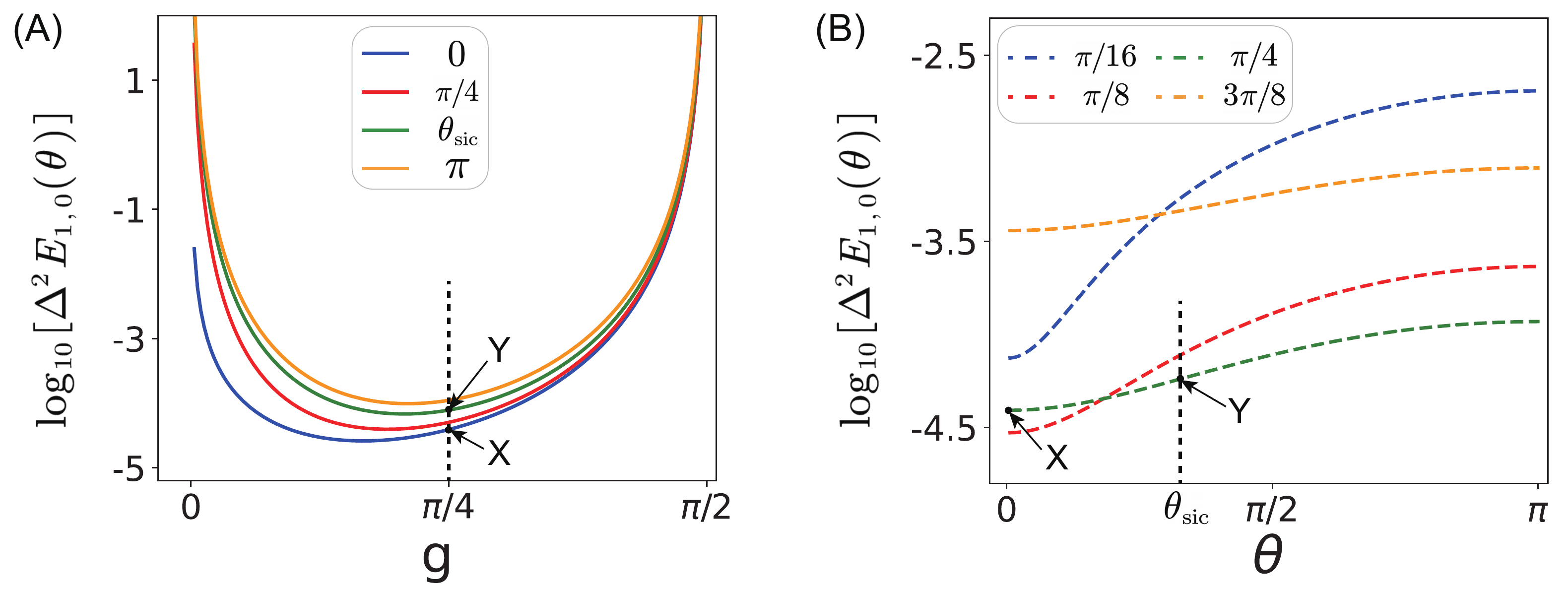}
\end{tabular}
\end{center}
\caption 
{ \label{Fig:prect}
The measurement precision of measuring the off-diagonal matrix entry $E_{1,0}(\theta)$ of the measurement operator $\hat{\Pi}(\theta)$ in two-dimensional quantum system. (A) The variance of $E_{1,0}(\theta)$ is plotted with different $g$ for four values of $\theta=0,\pi/4,\theta_{\text{sic}},\pi$ with $\theta_{\text{sic}} = \text{acos} (1/\sqrt{3})$. (B) The variance of $E_{1,0}(\theta)$ changes with different parameter $\theta$ for the coupling strength $g = \pi/16,\pi/8,\pi/4,3\pi/8$. Here, we take $\eta=1/2$ and $N = 12790$ to coincide with our experimental conditions. The points X and Y refer to the precision of directly measuring the off-diagonal matrix entry of the two-dimensional symmetric informationally complete positive operator-valued measure with the coupling strength $g=\pi/4$.
} 
\end{figure}

As a demonstration, we theoretically derive the precision of directly measuring the off-diagonal matrix entry $E_{1,0}(\theta)$ of a general measurement operator for a two-dimensional QS
\begin{eqnarray}
\hat{\Pi}(\theta) =\eta \left(
\begin{array}{ll}
\cos^2\theta  &  E_{0,1}(\theta)  \\
E_{1,0}(\theta)  &  \sin^2\theta
\end{array}
\right)
\end{eqnarray}
with different coupling strength $g$. According to the Eq. \eqref{Prec_err}, the variance of the off-diagonal matrix entry $E_{1,0}(\theta)$ is given by
\begin{equation}
\Delta^2 E_{1,0}(\theta) = \frac{(\sin^2\theta+ \sin^2 g)(1+2\sin^2 g)}{\eta N\sin^4(2g)}.
\end{equation}
In Fig. \ref{Fig:prect} (A), we show how the variance of $E_{1,0}(\theta)$ changes with different $g$ for four values of $\theta$. We find that the statistical errors $\Delta^2 E_{1,0}(\theta)$ become large with a small coupling strength ($g\rightarrow 0$ or $g\rightarrow \pi/2$), while the strong coupling strength ($g\rightarrow \pi/4$) significantly decreases the variance to $\Delta^2 E_{1,0}(\theta)_{|g=\pi/4} = (1+2\sin^2\theta)/(\eta N)$. We also compare the characterization precision of $E_{1,0}(\theta)$ associated with different POVM parameter $\theta$ in Fig. \ref{Fig:prect} (B). The statistical errors $\Delta^2 E_{1,0}(\theta)$ remain finite over all $\theta$ indicating that our protocol is applicable to characterize the arbitrary POVM of two-dimensional quantum system. In addition, the variance $\Delta^2 E_{1,0}(\theta)$ is related to the parameter $\theta$ but does not depend on the value of $E_{1,0}(\theta)$. This implies that the change of the off-diagonal matrix entries of the measurement operator, such as the dephasing and the phase rotation process will not affect the characterization precision. We note that the choice of the sequential observables $\hat{O}_B$ and $\hat{O}_A^{(k)}$ is indeed not unique. How to choose the optimal observables of the quantum system to achieve the best characterization precision remains an open question in the field of direct tomography. If the sequential observables of the quantum system are changed, the collective observables $\hat{R}_{B,A}$ and $\hat{T}_{B,A}$ of the meter states should also be changed correspondingly to reveal the matrix entries $E_{a_ja_k}^{(l)}$.

It has been shown that the completeness condition of the POVM $\{\hat{\Pi}_l\}$, $i.e.$, $\sum_l\hat{\Pi}_l=\hat{I}$, can be used to improve the precision of direct quantum detector tomography \cite{xu2020direct}. In the following, we prove that the same condition is also helpful to improve the precision in the direct characterization of $E_{a_ja_k}^{(l)}(j\neq k)$. Since the real part of the entries $E_{a_ja_k}^{(l)}$ satisfy $\sum_l \text{Re}[E_{a_ja_k}^{(l)}] = 0$, the value of $\text{Re}[E_{a_ja_k}^{(l)}]$ can be not only obtained by the direct measurement but also inferred from the entries of other POVM elements $\text{Re}[E_{a_ja_k}^{(u)}]$ ($u\neq l$) by $\text{Re}[E_{a_ja_k}^{(l)\circ}] = -\sum_{u\neq l}\text{Re}[E_{a_ja_k}^{(u)}]$. The extra information obtained by $\text{Re}[E_{a_ja_k}^{(l)\circ}]$ can be used to improve the measurement precision. To acquire the best precision, we adopt the weighted average of $\text{Re}[E_{a_ja_k}^{(l)}]$ and $\text{Re}[E_{a_ja_k}^{(l)\circ}]$ with the weighting factors $w$ and $w^\circ$, respectively. The optimal weighting factors satisfy the condition
\begin{eqnarray}
w &+& w^\circ = 1, {}\nonumber\\
w &\propto & \frac{1}{\Delta^2\text{Re}[E_{a_ja_k}^{(l)}]]}, {}\nonumber\\
w^\circ &\propto & \frac{1}{\sum_{u\neq l}\Delta^2\text{Re}[E_{a_ja_k}^{(u)}]},
\end{eqnarray}
leading to the optimal precision $\Delta^2\text{Re}[E_{a_ja_k}^{(l)\prime}] = ww^\circ /\sum_u \Delta^2 \text{Re}[E_{a_ja_k}^{(l)}]<\Delta^2\text{Re}[E_{a_ja_k}^{(l)}]$.

\section{Experiment}
In the experiment, we apply the DT protocol to characterize the symmetric informationally complete (SIC) POVM in the polarization degree of freedom (DOF) of photons. Since the coherence between two polarization base states only changes the off-diagonal entries of the measurement operators, we demonstrate that the dephasing and the phase rotation of the SIC POVM can be monitored by only characterizing the corresponding matrix entries.

\begin{figure}
\begin{center}
\begin{tabular}{c}
\includegraphics[height=6.7cm]{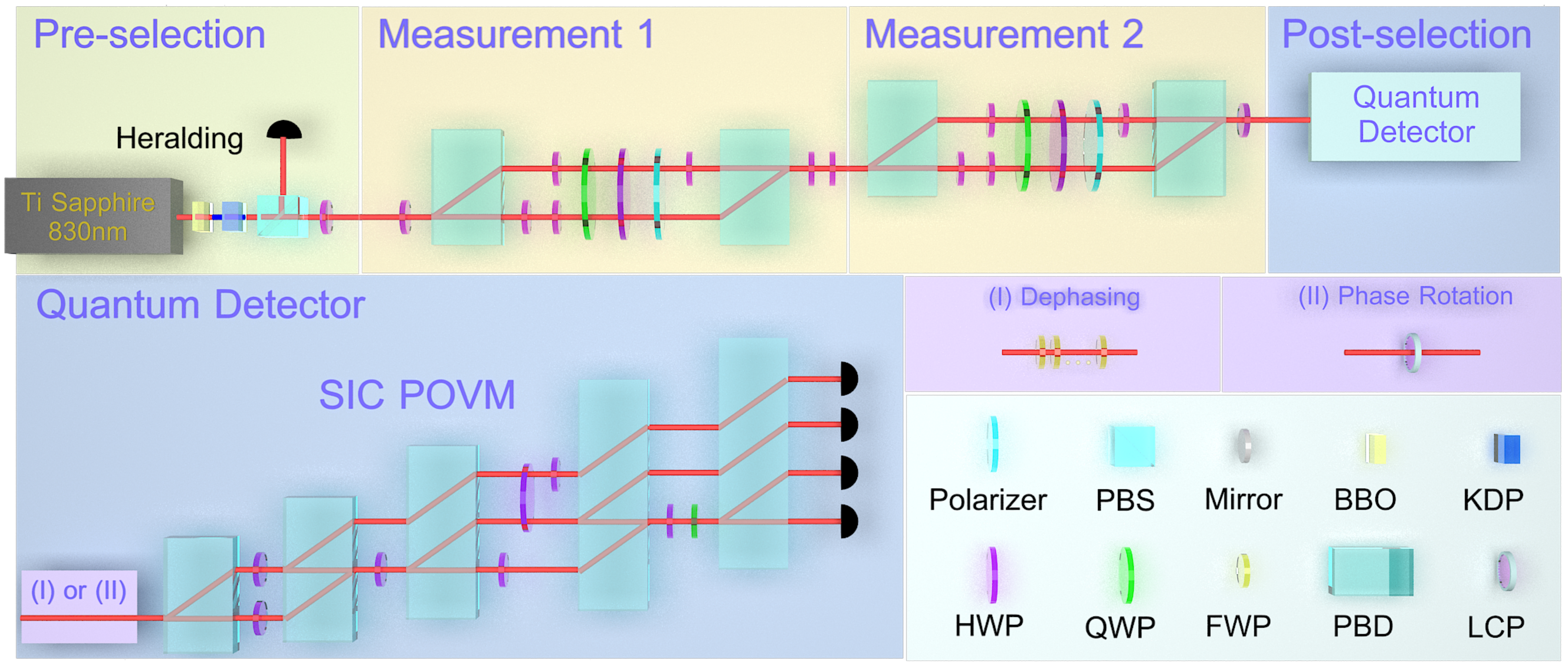}
\end{tabular}
\end{center}
\caption 
{ \label{expsetup}
The experimental setup of directly characterizing the evolution of the quantum measurement. The pulse laser at 830 nm inputs a BBO crystal for the up conversion. The generated photons at 415 nm gets through a KDP crystal for the spontaneous parametric down conversion, which simultaneously produces a pair of photons. The single photon is heralded by detecting the other one of the pair. The `Measurement 1' and `Measurement 2' modules successively implement the unitary transformation $\hat{U}_B$ and $\hat{U}_A^{(k)}$ as well as the joint measurement on the meter states. In the following, the unknown quantum detector performs the post-selection measurement on the polarization degree of freedom of photons. The quantum detector is composed of the operation of polarization evolution, $i.e.,$ `(\uppercase\expandafter{\romannumeral1}) Dephasing' and `(\uppercase\expandafter{\romannumeral2}) Phase Rotation' and the symmetric informationally complete positive operator-valued measure (SIC POVM) realized by the quantum walk. The abbreviations of the equipments are given as follows: PBS, polarizing beam splitter; BBO, $\beta$-barium borate crystal; KDP, potassium di-hydrogen phosphate; HWP, half wave plate; QWP, quarter wave plate; PBD, polarizing beam displacer; FWP, full wave plate; LCP, liquid-crystal plate.}
\end{figure}

The experimental setup is shown in Fig. \ref{expsetup}. We refer to the polarization DOF of photons as the QS with the eigenstates $|H\rangle$ and $|V\rangle$. Single photons generated by the spontaneous parametric down conversion pass through the polarizing beam splitter (PBS) and a half-wave plate (HWP) at $45^\circ$ to pre-select the QS to $|V\rangle$. The `Measurement 1' and `Measurement 2' modules implement the measurement of the observables $\hat{O}_B = |D\rangle\langle D| - |A\rangle\langle A|$ and $\hat{O}_A = |H\rangle\langle H| - |V\rangle\langle V|$, where $|D\rangle = (|H\rangle + |V\rangle)/\sqrt{2}$ and $|A\rangle = (|H\rangle - |V\rangle)/\sqrt{2}$. 

Here, we take the `Measurement 1' as an example to describe the working principle of the coupling scenario. The HWP at $22.5^\circ$ before the polarizing beam displacer (PBD) transforms the measurement basis $\{|D\rangle, |A\rangle\}$ into $\{|H\rangle, |V\rangle\}$ and the observable $\hat{\sigma}_z = |H\rangle\langle H|-|V\rangle\langle V|$ is measured between the two PBDs. The first PBD converts the DOF of the QS into the optical path with $|H\rangle\rightarrow |0\rangle$ and $|V\rangle\rightarrow |1\rangle$. The polarization of photons in each path initialized to $|H\rangle$ is used as the MS. Two HWPs arranged in parallel each on different path are respectively rotated to $g/2$ and $-g/2$ to realize the coupling between the QS and the MS. Afterwards, we measure the polarization of photons to extract the information of the MS by a quarter-wave plate (QWP), a HWP and a polarizer. The photons in two paths that pass through the polarizer recombine at the second PBD and the subsequent two HWPs at $45^\circ$ and $22.5^\circ$ recover the measurement basis to $\{|H\rangle, |V\rangle\}$. A similar setup of `Measurement 2' performs the measurement of the operator $\hat{O}_A$. Finally, the photons input the unknown detector for the post-selection. By collecting the photons that arrive the outputs, we obtain the measurement results.

We construct the SIC POVM $\{\hat{\Pi}_l\}$ with $\hat{\Pi}_l = \frac{1}{2}|\psi_l\rangle \langle \psi_l| (l=1,2,3,4)$ and
\begin{eqnarray}
|\psi_1\rangle &=& |H\rangle,{}\nonumber\\
|\psi_2\rangle &=& (|H\rangle - \sqrt{2}|V\rangle)/\sqrt{3},{}\nonumber\\
|\psi_3\rangle &=& (|H\rangle + \sqrt{2}e^{-i2\pi/3}|V\rangle)/\sqrt{3},{}\nonumber\\
|\psi_4\rangle &=& (|H\rangle + \sqrt{2}e^{i2\pi/3}|V\rangle)/\sqrt{3}
\end{eqnarray}
through the quantum walk to perform the post-selection measurement of the QS \cite{PhysRevLett.114.203602}. The dephasing of the POVM is realized by several full-wave plates (FWPs) which separate the wave packets in polarization states $|H\rangle$ and $|V\rangle$, i.e., $|\varphi(t_H)\rangle$ and $|\varphi(t_V)\rangle$ in the temporal DOF. This separation causes the dephasing of the POVM and the off-diagonal entries $E_{VH}^{(l)}$ are transformed to $E_{VH}^{(l),\text{D}} = E_{VH}^{(l)}\xi$ with the coefficient $\xi = \langle \varphi(t_H)|\varphi(t_V)\rangle$. The derivation of the dephasing process and the calibration of the coefficient $\xi$ are provided in the Appendix \ref{Sec:Appendix}. The phase rotation is implemented by the liquid crystal plate (LCP), which imposes a relative phase $\phi_{lc}$ between $|H\rangle$ and $|V\rangle$. The operation is equivalent to the unitary evolution $\hat{U}_{lc} = \exp(i\frac{\phi_{lc}}{2}\hat{C})$ of the input state, with $\hat{C} = |H\rangle\langle H| - |V\rangle\langle V|$. When the evolution is inversely performed on the SIC POVM, the nondiagonal elements $E_{VH}^{(l)}$ is transformed to $E_{VH}^{(l),\text{R}} = E_{VH}^{(l)}\exp(-i\phi_{lc})$. The calibration results of the $\phi_{lc}$ are shown in the Appendix \ref{subsec:A2}.

\begin{figure}[ht]
\begin{center}
\begin{tabular}{c}
\includegraphics[height=11cm]{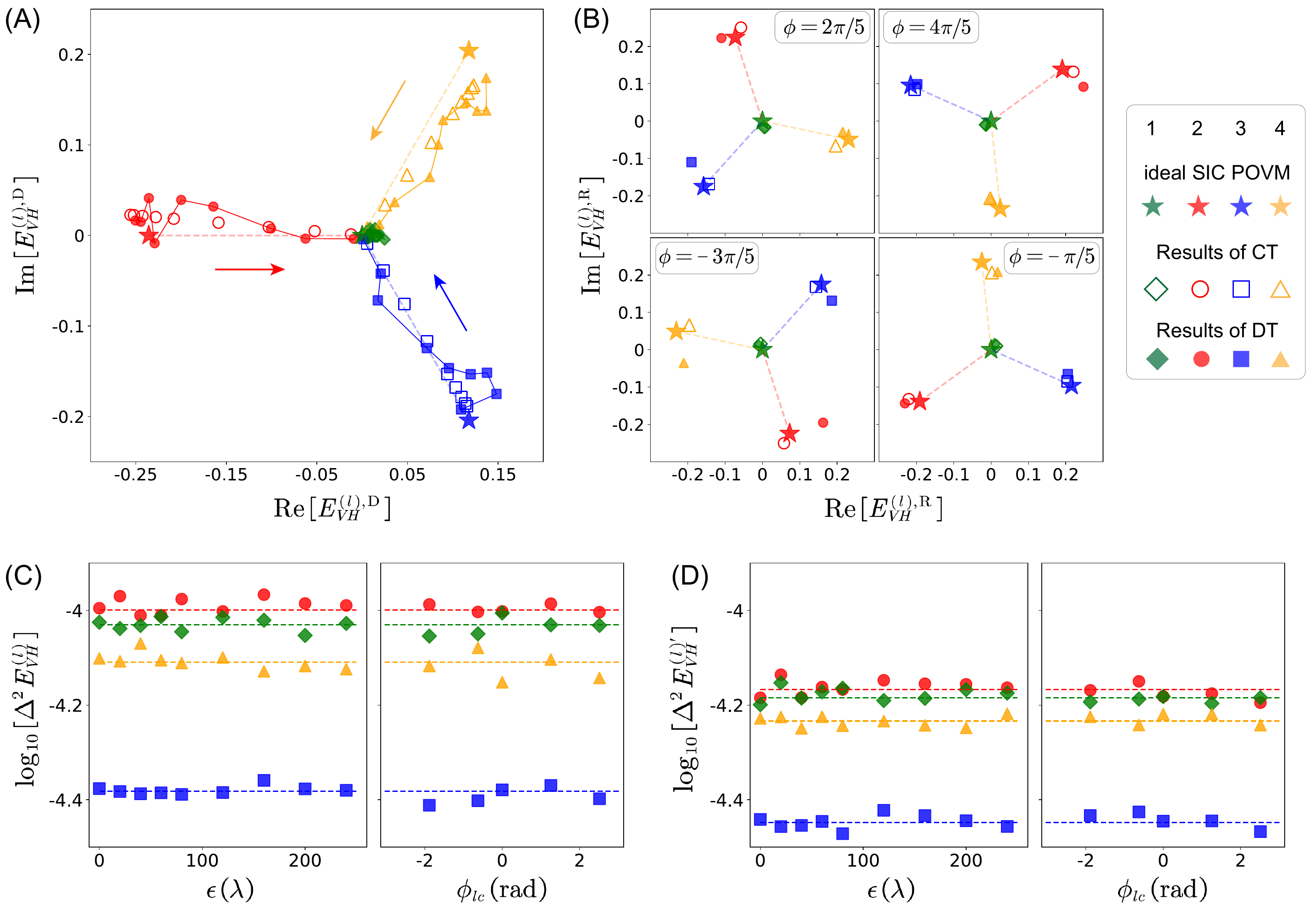}
\end{tabular}
\end{center}
\caption 
{ \label{expresults}
The experimental results and the corresponding statistical errors are illustrated. The real and the imaginary parts of the matrix entries $E_{VH}^{(l)}$ are plotted during the dephasing ($E_{VH}^{(l),\text{D}}$) and the phase rotation ($E_{VH}^{(l),\text{R}}$) of the polarization in (A) and (B), respectively. The results of the ideal symmetric informationally complete (SIC) positive-operator-valued measure (POVM), the conventional tomography (CT) and the direct tomography (DT) are represented by the pentagram, hollow markers and solid markers, respectively. In (A) and (B), we connect each pentagram with the point (0,0) indicating the evolution path of the ideal SIC POVM during the dephasing process as well as changes of the azimuth angles during the phase rotation process. The statistical errors of the matrix entries $E_{VH}^{(l)}$ are provided in (C) for both the dephasing and the phase rotation process. In (D), we illustrate the precision of $E_{VH}^{(l)\prime}$ after using the completeness condition of the POVM. The theoretical precision represented by the dashed lines in (C) and (D), is inferred from the experimental results of CT. The average photon number per unit time for one collective measurement of the meter states is about $N=12790$.
 }
\end{figure}

\section{Results}
\label{sect:results}
In Fig. \ref{expresults}, we compare the experimental results of DT with those of the conventional tomography (CT) as well as the ideal SIC POVM during the dephasing and phase rotation process. The detailed information of characterizing the experimental SIC POVM by CT is provided in the Supplementary Materials. The results of CT shown in Fig. \ref{expresults} are inferred from the experimental SIC POVM and the calibrated coefficient $\xi$ (during the dephasing process) or the phase $\phi_{lc}$ (during the phase rotation process). As shown in Fig. \ref{expresults} (A), the points in each connecting solid line along the direction of arrows correspond to the relative time delay $\epsilon = 0,20,40,60,80,120,160,200,240$ ($\lambda$). The increase of the relative time delay $\epsilon$ between the separated wave packets reduces the overlap of the temporal wavefunction $\xi =\langle \varphi(t_H)|\varphi(t_V)\rangle $, which leads to the dephasing of the quantum measurement. The relation between the relative time delay $\epsilon$ and the coefficient $\xi$ is calibrated in Fig. \ref{Calibration} (B) of Appendix \ref{subsec:A2}. Correspondingly, the modulus of $E_{VH}^{(l),D}$ gradually approaches to 0, implying that the quantum measurement becomes incoherent, $i.e.$, loses the ability of detecting the coherence information of a quantum state.

In Fig. \ref{expresults} (B), we plot $E_{VH}^{(l),R}$ during the phase-rotation process. The imposed voltages on the LCP is adjusted to obtain $\phi_{lc} = 2\pi/5$ and $4\pi/5$. A HWP at $0^\circ$ is placed before the LCP to obtain $\phi_{lc} = -3\pi/5$ and $-\pi/5$. The rotated points representing $E_{VH}^{(l),R}$ in the coordinate of its real and imaginary part indicates the phase rotation of the quantum measurement. During the phase rotation process, the modulus of $E_{VH}^{(l),R}$ remains unchanged, which indicates that the coherence of the quantum measurement maintains.

The total noise in the experiment contains the statistical noise and the technical noise. The statistical noise originates from the fluctuations of the input photon numbers per unit time due to the probabilistic generation of single photons, the loss in the channel and the finite trials of the experiment. The technical noise is caused by the experimental imperfections, e.g., the equipment vibration or the air turbulence. As shown in Fig. \ref{expresults} (A) and (B), the experimental results fluctuate around the theoretical predictions due to both the statistical noise and the technical noise. The technical noise can be reduced by isolating the noise source or adopting appropriate signal modulation. The statistical noise determines the ultimate precision that can be achieved for a specific amount of input resources, which is an important metric to evaluate whether a measurement protocol is efficient or not.

The statistical errors of the experimental results are shown in Fig. \ref{expresults} (C). The theoretical precision, represented by dashed lines in (C) and (D), is inferred by assuming that the matrix entries $E_{VH}^{(l)}$ of the experimental SIC POVM obtained by the CT are directly characterized. As a comparison, we can refer to Fig. \ref{Fig:prect} for the theoretical precision of the ideal SIC POVM, represented by the points $X$ ($\theta=0,g=\pi/4$) and $Y$ ($\theta=\text{acos}(1/\sqrt{3}),g=\pi/4$). Since the experimental SIC POVM deviates from the ideal SIC POVM, the precision of $l=2,3,4$ do not equate with each other. The experimental precision is obtained from the Monte Carlo simulation based on the experimental probability distribution and the practical photon statistics to eliminate the effect of the technical noise. Our results closely follow the theoretical predictions indicating that the precision of measuring the off-diagonal matrix entries of the POVM is immune to the dephasing and phase rotation of the quantum measurement. We can also find that the characterization precision after using the completeness condition in Fig. 4 (D) is significantly improved compared to the original precision in Fig. 4 (C).

\section{Discussion and Conclusions}
We have proposed a protocol to directly characterize the individual matrix entries of the general POVM, extending the scope of the direct tomography scheme. Our expression is rigorous for the arbitrary coupling strength, which allows to change the coupling strength to improve the precision and simultaneously maintain the accuracy. The statistical errors are finite over all the choice of the POVM parameter demonstrating the feasibility of our protocol for the arbitrary POVM. In particuliar, if the completeness condition of the POVM is appropriately used, the measurement precision can be further improved. Our results indicate that the characterization precision is not affected by the dephasing and phase rotation that only change the off-diagonal matrix entries of the measurement operators. Another typical noise is the phase diffusion meaning that the phase of the quantum measurements randomly jitters. According to the derivations in the paper \cite{brivio2010experimental}, the phase diffusion decreases the modulus of the off-diagonal matrix entries in a similar way to the dephasing in our work. Therefore, it is expected that the precision of our protocol is immune to the incoherent noise, such as phase diffusion.

Since some properties of quantum measurements may depend on a part of matrix entries of the measurement operators, this protocol allows us to reveal these properties without the full tomography. We experimentally demonstrate that the evolution of the coherence of a quantum measurement can be monitored through determining the off-diagonal matrix entries of the measurement operators. Our scheme makes no assumptions about the basis to represent the measurement operators. The choice of the basis depends on the specific conditions or can be optimized according to the research goals. For example, the quantum properties can be basis-dependent ($e.g.$, coherence), or are better revealed with proper choice of the basis ($e.g.$, entanglement). Our scheme provides the flexibility to characterize the matrix entries of the measurement operators in any basis of interest by adjusting the initial quantum state as well as the sequential observables while other parts of the theoretical framework remain unchanged. This feature is an advantage for us to explore the coherence properties or to seek the optimal entanglement witness \cite{horodecki2009quantum}.

Our protocol can be extended to high-dimensional quantum system, in which the coherence information of the quantum measurement among specified base states is of interest. The conventional QDT typically requires $d^2$ informationally complete probe states chosen from at least $d+1$ basis to globally reconstruct the POVM in $d$ dimensional quantum system. Thus, as the dimension $d$ increases, the preparation of the probe states becomes an experimental challenge and the computational complexity of the reconstruction algorithm is significantly increased. Both factors complicate the task of QDT for the high-dimensional quantum systems. In our scheme, the preparation of the initial states and the sequentially measured observables $\hat{O}_B$ and $\hat{O}_A^{(k)}$ are simply involved in two basis, i.e., the representation basis $\{|a_j\rangle\}$ and its Fourier conjugate $\{|b\rangle\}$. The matrix entries of the POVM can be directly inferred from the measurement results of the final meter states without resort to the reconstruction algorithm. When the matrix entries are sparse in the measurement operators, our scheme can further simplify the characterization process. Therefore, the direct protocol also shows potential advantages over the conventional QDT in completely determining the POVM due to its better generalization to high-dimensional quantum systems. In conclusion, by proposing a framework to directly and precisely measure the arbitrary single matrix entry of the measurement operators, our results pave the way for both fully characterizing the quantum measurement and investigating the quantum properties of it.

\appendix

\section{Dephasing and phase rotation of quantum measurements}
\label{Sec:Appendix}
\subsection{Theoretical derivation}
\label{subsec:A1}
A general POVM can be implemented through quantum walk with the unitary evolution $\hat{U}$ of the QS at the position $x=0$. After the quantum walk, the position $x=l$ corresponds to the POVM element
\begin{equation}
\hat{\Pi}_l = \text{Tr}_W\big[(|0\rangle\langle 0|\otimes \hat{I}) \hat{U}^\dagger (|l\rangle\langle l|\otimes \hat{I})\hat{U} \big],
\end{equation}
where $\text{Tr}_W\{\cdot\}$ denotes the partial trace in the walker position DOF. We implement the dephasing of the POVM $\{\hat{\Pi}_l\}$ by coupling the QS to the environment state $\rho_E$ under the Hamiltonian $\hat{H}_{SE} = \frac{\epsilon}{2} \delta(t-t_0)\hat{C}\hat{\Omega}$, in which $\hat{C}=|a_j\rangle\langle a_j|-|a_k\rangle\langle a_k|$ and $\hat{\Omega}$ are the observables of the QS and the environment, respectively. By reducing the environment DOF, the measurement operator $\hat{\Pi}_l$ is transformed to $\hat{\Pi}_l^D = \text{Tr}_E\big[\hat{U}^\dagger_{SE} \hat{\Pi}_l\otimes \rho_E \hat{U}_{SE} \big]$. We can infer that the dephasing process only changes the related matrix entries $E_{a_ja_k}^{(l)}$ to $E_{a_ja_k}^{(l),D} = E_{a_ja_k}^{(l)}\xi$ with the coefficient $\xi = \text{Tr}[\exp(-i\frac{\epsilon}{2}\hat{\Omega})\rho_E\exp(-i\frac{\epsilon}{2}\hat{\Omega})]$.

\subsection{Experimental Calibration}
\label{subsec:A2}
\begin{figure}[ht]
\begin{center}
\begin{tabular}{c}
\includegraphics[height=3.5cm]{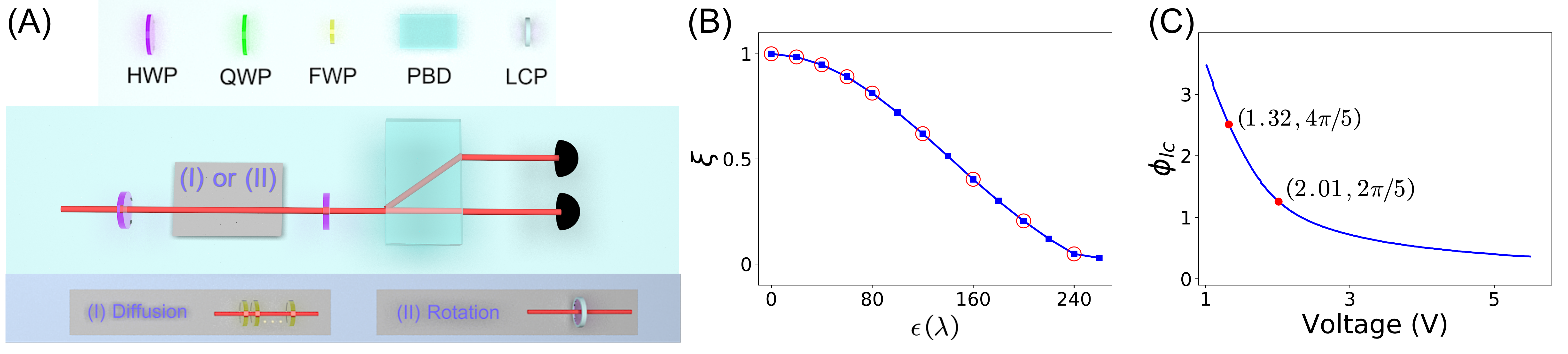}
\end{tabular}
\end{center}
\caption 
{ \label{Calibration}
The calibration of the equipment in the dephasing and the phase rotation process. (A) The calibration setup. (B) The coefficient $\xi$ changes with the time delay $\epsilon$ between the wave packets in states $|H\rangle$ and $|V\rangle$. (C) The relative phase $\phi_{lc}$ between the states $|H\rangle$ and $|V\rangle$ changes with imposed voltage.
 }
\end{figure}

To calibrate the relation between the coefficient $\xi$ and the relative time delay $\epsilon = |t_H-t_V|$, we construct the setup shown in Fig. \ref{Calibration} (a), in which both the half-wave plates (HWPs) are set to $22.5^\circ$. The photons in $|H\rangle$ inputs the calibration setup resulting in the final state after the second HWP 
\begin{equation}
\rho^{\text{D}} = \frac{1+\xi}{2}|H\rangle\langle H| + \frac{1-\xi}{2}|V\rangle\langle V|.
\end{equation}
Then, $\rho^{\text{D}}$ is projected to the basis $\{|H\rangle,|V\rangle\}$ with a polarizing beam displacer (PBD), obtaining the probabilities $P_H$ and $P_V$. The parameter $\xi$ is given by $\xi = P_H - P_V$. The relation between $\xi$ and the relative time delay $\epsilon$ is shown in the Fig. \ref{Calibration} (B), in which we take $\epsilon$ from 0 to 260 times the wavelength ($\lambda = 830nm$)  and the red circled points are adopted for the experiment.

The liquid crystal imposes a relative phase $\phi_{lc}$ between $|H\rangle$ and $|V\rangle$ controlled by the voltage. Through the calibration setup in Fig. \ref{Calibration} (A), the phase can be obtained by $\phi_{lc} = \arccos[2(P_H - P_V)]$. The calibration results of the relation between the phase $\phi_{lc}$ and the applied voltage are shown in Fig. \ref{Calibration} (C). Here, we adjust the voltages to 1.32V and 2.01V and the relative phases are approximately $4\pi/5$ and $2\pi/5$.

% \disclosures 
\subsection*{Disclosures}
The authors declare no conflicts of interest.
$^\dagger$These authors contribute equally to this work.

\subsection* {Acknowledgments}
This work was supported by the National Key Research and Development Program of China (Grant Nos. 2017YFA0303703 and 2018YFA030602) and the National Natural Science Foundation of China (Grant Nos. 91836303, 61975077, 61490711 and 11690032) and Fundamental Research Funds for the Central
Universities (Grant No. 020214380068).

\subsection* {Code, Data, and Materials Availability}
The computer software code, data are available by connecting to the corresponding authors.

%%%%% References %%%%%

\bibliographystyle{spiejour}   % makes bibtex use spiejour.bst

%%%%% Biographies of authors %%%%%

\vspace{2ex}\noindent\textbf{Liang Xu} is a postdoc fellow in Zhejiang Lab. He received his BS in physics of materials and PhD in Optics engineering from Nanjing University in 2014 and 2020, respectively. His current research interests include weak measurement, quantum metrology and quantum tomography.

\vspace{2ex}\noindent\textbf{Huichao Xu} is an assistant researcher at Nanjing University. He received his BS and MS degrees in Telecommunication from the University of Jiamusi and University of Liverpool in 2009 and 2011, respectively, and his PhD degree in quantun optics from the Nanjing University in 2020. His current research interests include quantum state generation and quantum detectors.

\vspace{2ex}\noindent\textbf{Lijian Zhang} is a professor at the College of Engineering and Applied Sciences in the Nanjing University. He received his BS and MS degrees in electrical engineering from the Peking University in 2000 and 2003, respectively, and his PhD degree in physics from the University of Oxford in 2009. He is the author of more than 50 journal papers and two book chapters. His current research interests include quantum optics and quantum information processing.
\vspace{1ex}
\noindent Biographies and photographs of the other authors are not available.

\listoffigures
\listoftables

\end{spacing}
\end{document}